\documentclass[twocolumn,showpacs,amsmath]{revtex4}

\usepackage{graphicx}%
\usepackage{here}

\newcommand\pictc[5]{\begin{figure}
                       \centerline{
                       \includegraphics[width=#1\columnwidth]{#3}}
		   \vspace{-0.1cm}
                   \protect\caption{\protect\label{fig:#4} #5 \vspace{-0.6cm}}
                    \end{figure}            }
\newcommand\pict[4][1.]{\pictc{#1}{!tb}{#2}{#3}{#4}}
\newcommand\rpict[1]{\ref{fig:#1}}

\newcounter{Fig}

\begin{document}
\begin{sloppy}

\title{Observation of transverse instabilities in optically-induced lattices}

\author{Dragomir Neshev} 
\author{Andrey A. Sukhorukov}
\author{Yuri S. Kivshar}
\affiliation{Nonlinear Physics Group, Research School of Physical Sciences and Engineering, Australian National University, Canberra ACT 0200, Australia}

\author{Wieslaw Krolikowski}
\affiliation{Laser Physics Center, Research School of Physical Sciences and Engineering, Australian National University, Canberra ACT 0200, Australia}

\begin{abstract}
We study experimentally the Bloch-wave instabilities in optically-induced lattices. We reveal two different instability scenarios associated with either the transverse modulational instability of a single Bloch wave, or the nonlinear inter-band coupling between different Bloch waves. These effects are compared with the transverse instability in homogeneous media, and it is shown that the periodic modulation of the refractive index {\em greatly enhances} the transverse instability.
\end{abstract}


\maketitle

Nonlinearity-induced instabilities are observed in many different branches of physics, and they provide probably the most dramatic manifestation of strongly nonlinear effects that can occur in Nature. Transverse or symmetry-breaking instabilities of solitary waves were predicted theoretically almost 30 years ago (see details in a review paper~\cite{Kivshar:2000-117:PRP}), but only recently both transverse and spatiotemporal instabilities were observed experimentally for different types of (bright and dark) spatial optical solitons~\cite{Kivshar:2003:OpticalSolitons}.

The effects of modulational and transverse instabilities have been observed mostly for continuous media, whereas theoretical results on nonlinear wave dynamics in Bragg gratings and waveguide arrays indicate that periodicity can strongly influence such instabilities~\cite{Kivshar:2003:OpticalSolitons,Aceves:1994-1186:OL}. It was recently demonstrated that lattice solitons can form in an array of optically-induced waveguides inside a photorefractive crystal~\cite{Fleischer:2003-23902:PRL, Neshev:2003-710:OL}. When the lattice is created in one spatial direction, and the beams are elongated along the orthogonal transverse direction, lattice solitons closely resemble discrete solitons existing in a one-dimensional waveguide array~\cite{Silberberg}. However, this analogy is limited since elongated beams may exhibit {\em transverse instabilities}. In this Letter, we study experimentally nonlinear beam self-action in optically-induced lattices, and reveal novel instability mechanisms which are initiated due to the medium periodicity.

\pict[0.8]{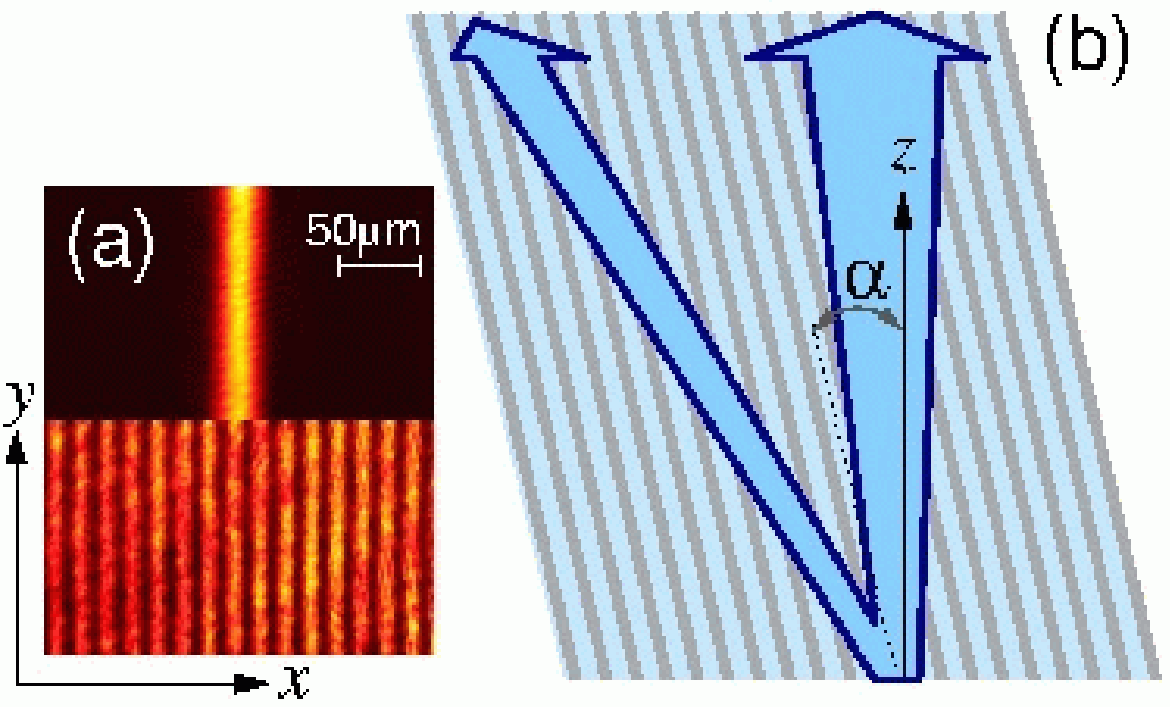}{excitation_scheme}{
(a) Input profiles of the stripe (top) and interference pattern (bottom). (b) Excitation scheme for observation of the inter-band coupling and transverse instability.}

We create a periodic lattice by interfering two ordinary polarized laser beams in a photorefractive Strontium Barium Niobate (SBN:60) crystal, which is externally biased with a DC field along the $x$-axis (crystalline $c$-axis)~\cite{Fleischer:2003-23902:PRL, Neshev:2003-710:OL}. The grating is constant along the propagation direction $z$, due to a low material nonlinearity for ordinary polarized light~\cite{Fleischer:2003-23902:PRL}. The polarity of the biasing field is such that an extraordinary polarized probe beam experiences strong focusing nonlinearity. The beam also feels the periodic potential created by the interference pattern, which lead to the beam scattering similar to the light propagation in waveguide arrays~\cite{Silberberg}. 

The extraordinary polarized beam is focused by a cylindrical lens, thus forming a homogeneous bright stripe at the input face of the crystal oriented along the fringes of the interference pattern [Fig.~\rpict{excitation_scheme}(a)]. The initial full-width-half-maximum of the stripe is 25.0(7)$\mu$m. The stripe diffracts and scatters in $x$-direction as it propagates inside the periodically modulated crystal [Fig.~\rpict{excitation_scheme}(b)]. Its propagation is governed by the nonlinear Schr\"odinger equation with a periodic potential, which in the case of optically induced gratings can be approximately written as 
$i {\partial_z \psi} + \frac{1}{2}\triangle_{\perp} \psi - \gamma [I_d + I_0 \cos^2(\pi x/d)+ |\psi|^2]^{-1} \psi =0,$
where $\triangle_{\perp}$ is the transverse Laplacian, $I_d$ is the dark irradiance, $I_0$ is the maximum intensity of the grating, $\gamma$ is a nonlinear coefficient linearly dependent on the applied electric field, and $d$ is the lattice period. The propagation of the probe beam strongly depends on the incident angle $\alpha$, due to the band-gap structure of wave spectrum of the induced periodic structure. At the input, the beam excites a superposition of Bloch waves, which are found as the linear eigenmodes of the periodic structure satisfying the symmetry relation
%
   $\psi( x, y, z ) = \psi( x + d, y, z ) \exp(i K_b x /d + i \beta z),$
%
where $K_b$ is the Bloch wavenumber, and $\beta$ is the propagation constant of the Bloch wave. By changing the intensity of the grating, the voltage on the crystal and/or the period of the interference pattern, we can change the induced periodic potential and subsequently modify the internal profile $\psi(x)$ of the excited Bloch waves. Dependence of the Bloch-wave spectrum of the beam inclination angle is characterized by analyzing the diffracted beam profile at the output face of the crystal~\cite{andrey_cleo} and experimentally we detect the output probe beam intensity distribution on a CCD camera.

We demonstrate below that the instability development in the vertical ($y$) direction strongly depends on the Bloch-wave spectrum. For this purpose we monitor the stripe profile with increasing of the laser power. To compare our results with the homogeneous case, we can remove the induced grating by making the two interfering beams mutually incoherent. This allows us to preserve the same amount of illumination inside the crystal and correspondingly the same strength of the nonlinearity. Details of the experimental arrangements are similar to those presented in Ref.~\cite{Neshev:2003-710:OL}.

Firstly, we investigate the nonlinear evolution for {\it collinear beam propagation} with respect to the induced lattice, when only Bloch waves from the first band are excited and Bragg scattering from the periodic structure is absent. In Fig.~\rpict{collinear_excitation} we show representative examples of our experimental results. The period of the grating for these experiments was $d=15.3\mu$m. The crystal was biased with electric field 4400~V/cm and the ratio between the grating intensity and the dark irradiance of the crystal was estimated to $I_{0}/I_{d}\sim 3$. When the grating is off, the extraordinary polarized probe beam focuses, and the stripe forms a quasi-one-dimensional soliton with homogeneous intensity along the $y$ direction [Fig.~\rpict{collinear_excitation}(a)]. When the grating is switched on, the beam forms a quasi-one-dimensional lattice (or discrete) soliton with a profile dependent on the initial horizontal position of the beam with respect to the lattice, as discussed in Ref.~\cite{Neshev:2003-710:OL}. However, the beam becomes spatially unstable along the transverse direction and starts to break [Fig.~\rpict{collinear_excitation}(b,c)]. The actual instability depends on the initial position of the beam with respect to the induced grating as can be seen from the comparison of Figs.~\rpict{collinear_excitation}(b) and (c).

We have detected several contributions toward the observed scenarios of the transverse instability. First, the perturbation in $x$ direction induced by the grating changes the conditions for instability in the vertical direction~\cite{Chen:2002-66601:PRE}, increasing the instability growth rate and causing the beam to break-up. The reason for this is simple. The beam is localized in the induced waveguides and it consists of several vertical stripes with narrower widths than the non-modulated beam width at the same laser power. Therefore, these thin stripes will tend to brake-up much faster than the wider non-modulated beam. Increasing the degree of power localization in an individual waveguide leads to growing instability of the Bloch wave~\cite{Kartashov:2003-1273:JOSAB}.
Second contribution comes from the inhomogeneities of the lattice-induced waveguides, which are not ideal in vertical direction. Such imperfections cause differences in the coupling between the neighboring waveguides along $y$ direction and therefore accelerate the development of the transverse instability. 

To compare the instability growth rate in the periodic lattice with that for the continuous system under the same input conditions, we switched off the grating and increased the laser-beam power until break-up of the stripe is observed Fig.~\rpict{collinear_excitation}(d). The striking result of this experiment is that {\em more than two orders of magnitude} higher beam power is needed in order to achieve qualitatively similar break-up of the stripe. For such high beam powers, the presence of the induced lattice leads to a full break-up of the signal beam as seen in Fig.~\rpict{collinear_excitation}(e).

\pict[0.99]{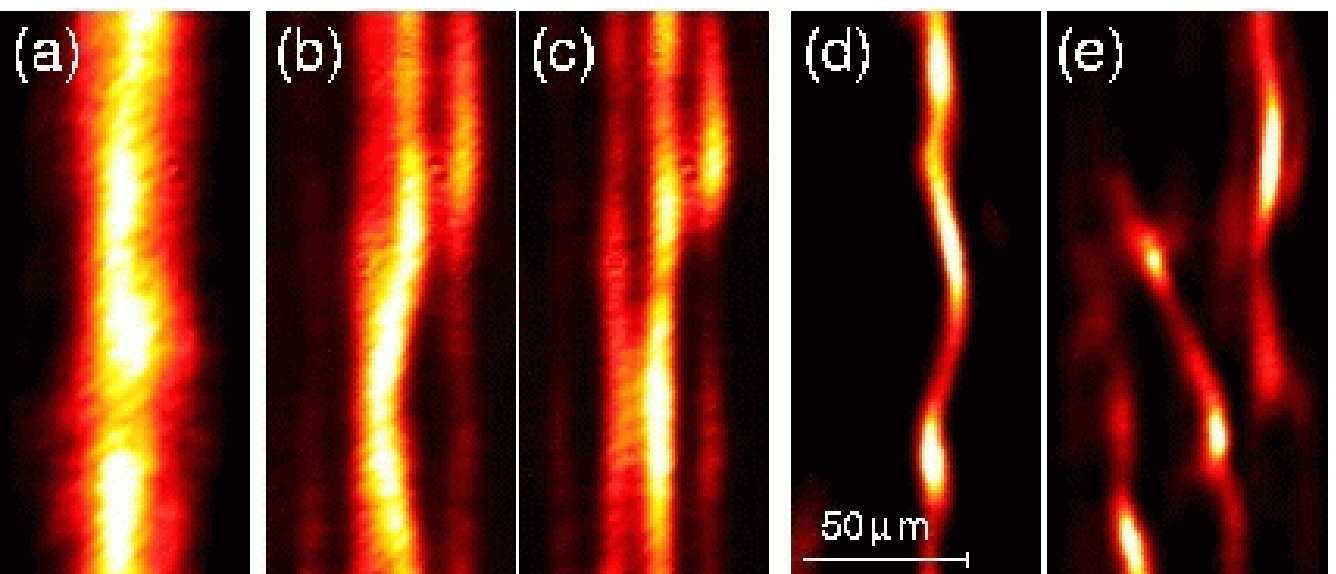}{collinear_excitation}{
Transverse beam instability under collinear excitation of the periodic lattice. (a) Uniform stripe when grating is off; (b,c) transverse break-up of the stripe due to the presence of the lattice for two different positions of the input beam (beam power 0.8$\mu$W), (d) transverse break-up of the stripe without lattice at high beam power (95$\mu$W). (e) full break-up of the stripe corresponding to (d) in the presence of the lattice.} 

We now investigate the development of the transverse instability which arises due to the {\it nonlinear interaction between two (or more) Bloch waves} as the fundamental modes of the lattice. To achieve interaction between two Bloch waves, we incline the grating with respect to the probe beam at an angle $\alpha=25$mrad, measured in air~\cite{experiment}, which is slightly less than twice the Bragg angle ($\alpha_{Br}=$13.5mrad). We induce a strong grating with a period $d=19.8\mu$m and ratio $I_{0}/I_{d}\sim 10$. In this case, the induced waveguides can support higher-order modes and the excited Bloch waves display a double-peak structure as seen from the plot in Fig.~\rpict{noncollinear_excitation}(a). The applied electric field on the crystal is 3600~V/cm. 

\pict[0.85]{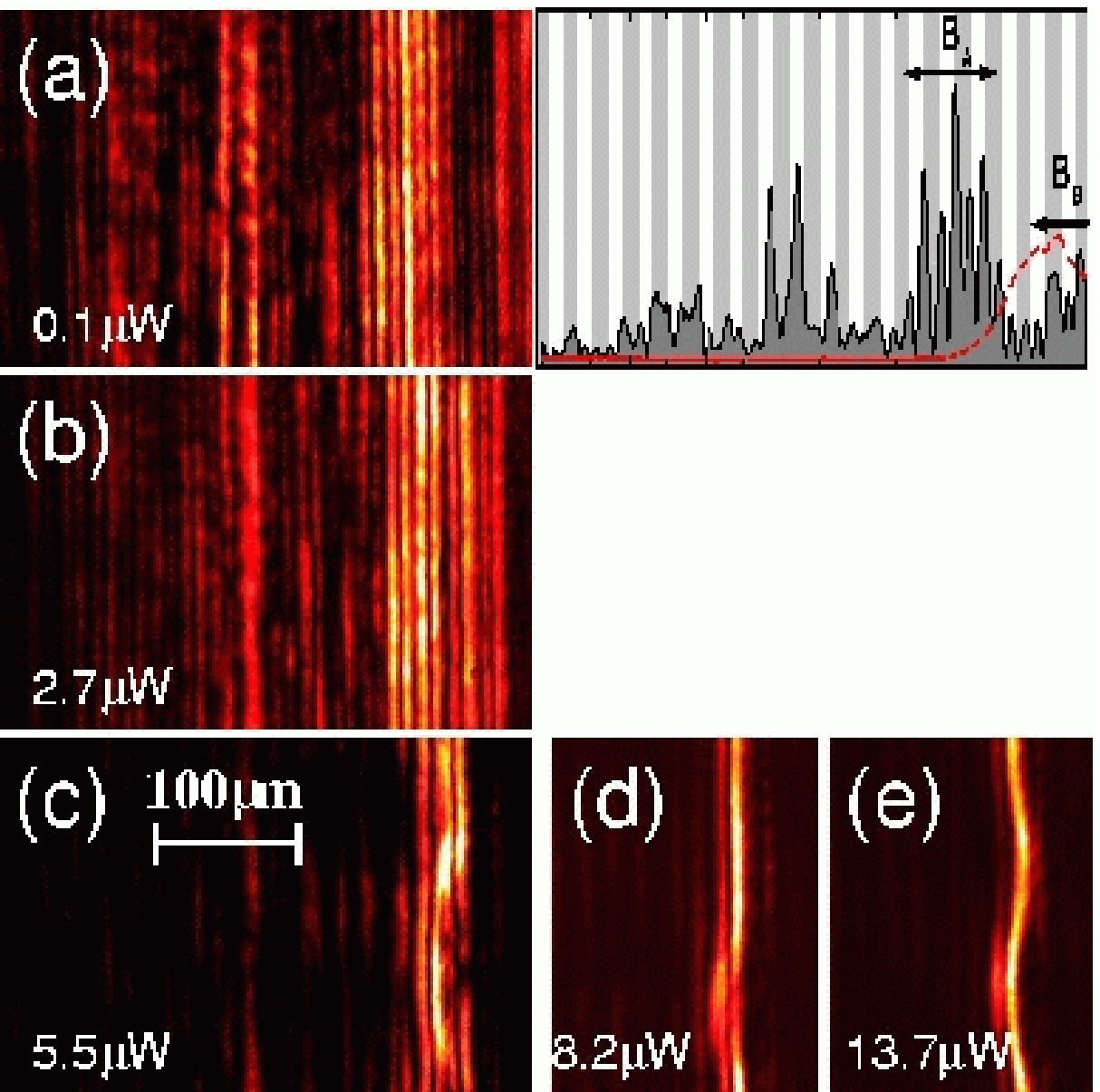}{noncollinear_excitation}{
Transverse instability under non-collinear excitation of the periodic lattice. (a) Linear excitation of different Bloch waves. The two interacting bands are marked in the plot as B$_{\rm A}$ and B$_{\rm B}$. The dash line represents the position of the beam without the grating. The shaded regions represent minima of the grating intensity. (b-e) Beam profiles at higher laser powers and the development of the transverse instability.}

A typical profile for the linear scattering is shown in Fig.~\rpict{noncollinear_excitation}(a). The dashed line shows the position of the beam when the grating is off. We select this particular regime in order to excite two Bloch waves B$_{\rm A}$ and B$_{\rm B}$ [denoted in Fig.~\rpict{noncollinear_excitation}(a)] next to each other and belonging to the first and the second excited spectral band. When the beam intensity is increased, the two bands start to move toward each other as seen in Fig.~\rpict{noncollinear_excitation}(a-c). This is due to the fact that the two Bloch waves have different dispersion relations and move into opposite directions at higher beam power~\cite{andrey_cleo}. The bands B$_{\rm A}$ and B$_{\rm B}$ start to mix together and strongly interact at about 5.5$\mu$W, as seen in Fig.~\rpict{noncollinear_excitation}(c). Due to the inter-band interaction the two Bloch waves exchange energy, which is a process strongly dependent on the intensity. Any small variation in the intensity distribution along the vertical $y$-axis is enhanced due to the inter-band energy exchange and causes the transverse break-up of the beam. 

To check that this effect is indeed due to the energy exchange between the two Bloch waves and not due to inhomogeneities in the induced waveguide, we have tested the instability dynamics (at 5.5$\mu$W probe beam power) as a function of the excitation position of the beam with respect to the grating. For this purpose we changed the phase of one of the interference beams, thus shifting the position of the maxima of the grating along $x$-axis for a full grating period. No significant difference in the breaking scenario was observed in this measurement. The break-up was found to be position insensitive in contrast to the collinear excitation.

At even higher intensities the inter-band mixing tends to disappear since one of the bands starts to dominate and at high enough laser powers only single band exists~\cite{andrey_cleo}. At these intensities the type of instability discussed earlier for the collinear excitation is present~[Fig.~\rpict{noncollinear_excitation}(d,e)]. It is comparable to the instability shown in Fig.~\rpict{collinear_excitation}(b,c). At much higher (about ten times) laser powers the beam starts to break in well defined circular spots and then transverse instability typical for a homogeneous medium is present. This is so because the probe beam-induced refractive index is dominant over the grating-induced index change.

We note here that the transverse instability due to the inter-band mixing is sensitive to the initial excitation. At different (lower) angles such scenario is not observed since there were no conditions under which two bands would strongly overlap and mix with each other. However, instability of a single band can always be observed at higher powers, similar to the examples shown in Fig.~\rpict{collinear_excitation}(b,c) and Fig.~\rpict{noncollinear_excitation}(d,e).
The amazing feature which occurs in the experimental realization presented in Fig.~\rpict{noncollinear_excitation} is that {\em the different instability mechanisms can be clearly distinguished} as they appear at different power levels. It allows us to identify them in our experiments.

At last, we investigated experimentally the influence of additional, systematic factors which facilitate the instability-induced breaking of the stripe. Such a factor could be a slight vertical tilt of the stripe with respect to the fringes of the grating. This is possible if the beam is not perfectly vertical along $y$-axis. We tested this possibility experimentally by intentionally tilting the stripe (correspondingly tilting of the cylindrical focusing lens). In this case, we consequently excite different lattice modes~\cite{Neshev:2003-710:OL} along $y$-axis, which leads to a correspondingly {\it regular} break-up of the stripe in contrast to the instability patterns presented above.

In conclusion, we have observed, for the first time to our knowledge, the development of transverse modulational instability of lattice soliton stripes and Bloch waves in optically-induced lattices. We have revealed two distinctive scenarios of the Bloch-wave instability associated with either the transverse instability of a single Bloch wave enhanced by the periodic potential, or the strong inter-band mixing of different Bloch waves.

The authors thank Elena Ostrovskaya and Zhigang Chen for valuable discussions. This work was partially supported by the Australian Research Council.

\vskip-0.5cm

\end{sloppy}

\begin{thebibliography}{}
\bibitem{experiment} 
Experimentally, the angle between the induced waveguides and the probe beam is measured in air by the degree of rotation of the beam splitter combining the orthogonally polarized beams. The actual angle inside the crystal is $n=2.3$ times smaller.
\end{thebibliography}
\end{document}